\documentstyle[preprint,aps,epsf]{revtex}
\begin{document}
\draft

\title{
\vspace{-3.0cm}
\begin{flushright}  
{\normalsize UTHEP-349}\\
\vspace{-0.3cm}
{\normalsize 1996 }\\
\end{flushright}
%
Chiral zero modes on the domain-wall model\\ in 4+1 dimensions
}

\author{ S. Aoki and K. Nagai}
\address{Institute of Physics,  University of Tsukuba,
         Tsukuba, Ibaraki 305, Japan}

\date{\today}
\maketitle

\begin{abstract}
We investigate an original domain-wall model 
in 4+1 dimensions numerically
in the presence of U(1) dynamical gauge field
only in an extra dimension,
corresponding to a weak coupling limit
of 4-dimensional physical gauge coupling.
Using a quenched approximation 
we carry out numerical simulation for this model 
at $\beta_{s} (= 1 / g^{2}_{s}) =$ 
0.29 (``symmetric'' phase) and
0.5 (``broken'' phase), 
where $g_s$ is the gauge coupling constant 
of the extra dimension.
In the broken phase,
we found that
there exists a critical value of a domain-wall mass $m_{0}^{c}$ 
which separates a region with a fermionic zero mode on the domain wall
from the one without it
in the same case of (2+1)-dimensional model.
On the other hand, in the symmetric phase,
our numerical data suggest
that the chiral zero modes disappear
in the infinite limit of 4-dimensional volume.
From these results it seems difficult to construct the U(1) lattice chiral gauge 
theory via an original domain-wall formulation.
\end{abstract}
\pacs{11.15Ha, 11.30Rd, 11.90.+t}

\narrowtext
\section{Introduction}
\label{sec:int}
Although perturbative aspects of electro-weak interaction
are well described by the standard model,
its non-perturbative phenomenon such as baryon number asymmetry
have to be investigated beyond the perturbation theory.
At present, the most powerful  non-perturbative technique 
is the lattice field theory.
However it is non-trivial to define the standard model on a lattice
since it is a type of chiral gauge theories,
construction of which is one of the long-standing problem
of lattice field theory:
Because of the fermion doubling problems, 
a naively $D$-dimensional discretized lattice fermion field 
yields $2^D$ fermion particles, 
half of one chirality and half of the other, 
so that the theory becomes non-chiral\cite{nielnino}.
Several lattice approaches have been proposed,
but so far none of them have been proven 
to work successfully\cite{latchiral}.

D. Kaplan has proposed a new construction of lattice chiral gauge 
theories via domain-wall models\cite{kaplan}.
Starting from a vector-like gauge theory in $2k+1$ dimensions
with a fermion mass term being a shape of a domain-wall 
in the (extra) $(2k+1)$th dimension,
he showed in the weak gauge coupling limit that 
a massless chiral state arises as a zero mode 
bound to the $2k$-dimensional domain-wall 
while all the doublers have large masses of the lattice cut-off scale.
It has been also shown that the model works well for smooth background gauge
fields\cite{aokihirose,aokihirose1,jansen,GJK}.

Two simplified variants of the original Kaplan's domain-wall model
have been proposed: an ``overlap formula''\cite{naraneu,overlap} and 
a ``waveguide model''\cite{waveg,waveg1}. 
Gauge fields appeared in these variants
are $2k$-dimensional and are independent of the extra $(2k+1)$th coordinate,
while those in the original model are $(2k+1)$-dimensional and depend on
the extra $(2k+1)$th coordinate. 
These variants work successfully for 
smooth background gauge fields
\cite{constchi1,constchi2,constchi3,constchi4,constchi5,aokilevin,shamir,constchi},
as the original one does. 
Non-perturbative investigations for these variants
seems easier than that for the original model 
due to the simpler structure of gauge fields.

However it has been reported\cite{waveg,waveg1} that
the waveguide model in the weak gauge coupling limit
cannot produce chiral zero modes needed
to construct chiral gauge theories.
In this limit, if gauge invariance were maintained,
pure gauge field configurations equivalent to the unity 
by gauge transformation would dominate 
and gauge fields would become smooth.
In the set-up of the waveguide model, however, $2k$-dimensional gauge
fields are non-zero only in the layers near domain wall (waveguide) ,
so that the gauge invariance is broken in the edge of the waveguide.
Therefore, even in the weak gauge coupling limit,
gauge fields are no more smooth and become very ``rough'',
due to the gauge degrees of freedom appeared to be dynamical in this edge.
As a result of the rough gauge dynamics,
a new chiral zero mode with the opposite chirality to the original zero mode 
on the domain wall appears in the edge, so that
the fermionic spectrum inside the waveguide becomes vector-like.
It has been claimed\cite{waveg,waveg1} that this ``rough gauge problem'' also
exists in the overlap formula 
since the gauge invariance is broken 
by the boundary condition at the infinity of the extra 
dimension\cite{aokilevin,shamir}. 
Furthermore an equivalence between the waveguide model 
and the overlap formula has been pointed out for
the special case\cite{gswgol}. 
Although the claimed equivalence 
has been challenged in Refs.\cite{overwave,constchi}, 
it is still crucial for the success of the overlap formula to solve
the ``rough gauge problem'' and to show the existence of a chiral zero mode
in the weak gauge coupling limit.

In the original model 
there are two inverse gauge couplings $\beta=1/g^2$ 
and $\beta_s=1/g_s^2$, 
where $g$ is the coupling constant in (physical) $2k$ dimensions 
and $g_s$ is the one in the (extra) $(2k+1)$th dimension.
Very little are known about this model except $\beta_s=0$ 
case\cite{waveg,aoitnioshi,kornip} where the spectrum seems vector-like
and the case of (2+1)-dimensional U(1) model\cite{aokinagai}.
Since perturbation theory 
for the physical gauge coupling $g$ is expected to hold,
the fermion spectrum of the model can be determined
in the limit that $g\rightarrow 0$.
In this weak coupling limit, 
all gauge fields in the physical dimensions can be gauged away, 
while the gauge field in the extra dimension is still dynamical 
and its dynamics is controlled by $\beta_s$. 
Instead of the gauge degrees of freedom
in the edge of the waveguide, 
$(2k+1)$th component of gauge fields represent 
roughness of $2k$-dimensional gauge fields.
An important question is whether the chiral zero mode on the domain wall
survives in the presence of this rough dynamics.
The dynamics of the gauge field in this limit is equivalent to
$2k$-dimensional scalar model with $2 L_s$ independent copies 
where $2 L_s$ is the number of sites in the extra dimension. 
In general at large $\beta_s$ 
such a system is in a ``broken'' phase 
where some global symmetry is spontaneously 
broken, while at small $\beta_s$ the system is in a ``symmetric'' phase.
Therefore there exists a critical point $\beta_s^c$, and it is likely
that the phase transition at $\beta_s=\beta_s^c$ is continuous 
(second or higher order).
The ``gauge field'' becomes rougher and rougher at smaller $\beta_s$.
Indeed we know that the zero mode disappears at $\beta_s=0$\cite{aoitnioshi},
while the zero mode exists at $\beta_s=\infty$ ( free case ).
So far we do not know the fate of the chiral zero mode in the intermediate
range of the coupling $\beta_s$. There are the following three possibilities:
(a) The chiral zero mode always exists except $\beta_s=0$. In this case
we may likely construct a lattice chiral gauge theory
in both broken ($\beta_s > \beta_s^c$) and symmetric ($\beta_s < \beta_s^c$),
and the continuum limits may be taken at $\beta_s = \beta_s^c$.
This is the best case for the domain-wall model.
(b) The chiral zero mode exists only in the broken phase 
($\beta_s > \beta_s^c$). 
In this case the domain-wall method
can describe a lattice chiral gauge theory in the broken phase 
at finite cut-off. However it is likely that
the continuum limit taken at $\beta_s = \beta_s^c$ from above
leads to a vector gauge theory.
(c) No chiral zero mode survives except $\beta_s=\infty$. The original
 model can not describe lattice chiral gauge theories at all.
It is very important to determine which possibility is indeed realized
in the domain-wall model.

Instead of  (4+1)-dimensional models,
we have recently investigated
a (2+1)-dimensional U(1) model\cite{aokinagai}.
Using a quenched approximation
we have carried out a numerical simulation
to see whether chiral zero modes exist or not in this model.
In the weak coupling limit of the physical gauge coupling
the (2+1)-dimensional U(1) gauge system 
is reduced to the 2-dimensional U(1) spin system 
{\footnote{This is explained in the next section}}.
Strictly speaking, 
there is no order parameter in the 2-dimensional U(1) spin system. 
On a large but finite lattice, however, 
the behavior of the model is similar 
to the one of a 4-dimensional scalar model:
On a finite lattice
we regard the Kosterlitz-Thouless phase as the ''symmetric'' phase
and the spin-wave phase as the ''broken'' phase.
In the ''broken'' phase
we have numerically found that 
there exists a critical value of a domain-wall mass $m_{0}^c$
which separates a region
with a fermionic zero mode on the domain wall
from one without it.
In the ''symmetric'' phase 
the critical values of the domain wall mass seems to also exist 
but is very close to its upper bound $m_0=1$,
so that the region with a fermionic zero mode is very narrow.
Because of the difficulty observed in the numerical simulation near $m_0=1$ 
we cannot exclude a possibility 
that the existence of the zero mode is an artifact of finite lattice size effects.
Further simulation we have made on larger lattice sizes
can not give a definite conclusion.
Since, as mentioned before, the phases of the model in the infinite volume 
limit is different from the ones of the 4-dimensional model,
we did not attempt to increase lattice sizes further, 
for example $100^2$, to see the fate of zero mode in the symmetric phase.
Instead we have decided to investigate the (4+1)-dimensional U(1) model 
directly, to obtain the definite conclusion 
on the existence of zero modes in the symmetric phase.

In this paper,
in order to know the fate of the chiral zero mode,
we have carried out a numerical simulation of a domain-wall model
in 4+1 dimensions
with a quenched U(1) gauge field in the $\beta=\infty$ limit.
In Sec.2 , 
we have defined our domain-wall model 
with dynamical gauge fields.
We have calculated a fermion propagator 
by using a kind of mean-field approximation,
to show that there is a critical value 
of the domain-wall mass parameter 
above which the zero mode exist.
The value of the critical mass may depend on $\beta_s$,
which controls the dynamics of the gauge field.
In Sec.3 , 
we have calculated the fermion spectrum numerically
using quenched approximation at $\beta_s = 0.29 , 0.5$
and at various values of domain-wall masses.
We have found that in the broken phase ($\beta_s = 0.5$) 
there exists the range of a domain-wall mass parameter
in which the chiral zero mode survives on the domain-wall.
In the symmetric phase ($\beta_s = 0.29$), however, 
from data on several lattice sizes
we have found an numerical evidence that
the chiral zero mode disappears 
in the infinite volume limit of 4-dimensional Euclidean space-time.
Our conclusions and some discussions are given in Sec. 4.
%
%
%
%
\section{domain-wall model}
\subsection{Definition of the model}
We consider a vector gauge theory in $D=2k+1$ dimensions
with a domain-wall mass term, 
which has a kink-like mass term
in the coordinate of an extra dimension.
This  domain-wall model is originally proposed by Kaplan\cite{kaplan},
and a fermionic part of the action 
is reformulated by Narayanan-Neuberger\cite{naraneu},
in terms of a $2k$-dimensional theory.
The model is defined by the action 
\begin{equation}
S = S_{G} + S_{F} , 
\end{equation}
where $S_{G}$ is the action of a dynamical gauge field , $S_{F}$ is the fermionic action.
$S_{G}$ is given by
\begin{equation}
S_{G} = \beta \sum_{n,\mu > \nu} \sum_{s} \left\{ 1 - {\rm Re Tr} \left[ 
U_{\mu \nu}(n,s) \right] \right\}
+ \beta_{s} \sum_{n,\mu} \sum_{s} \left\{ 1 - {\rm Re Tr} \left[ 
U_{\mu D}(n,s) \right] \right\} , 
\end{equation}
where $\mu ,  \nu$ run from $1$ to $2 k$ , 
$n$ is a $2 k$-dimensional lattice point , 
and $s$ is a coordinate of an extra dimension.
$U_{\mu \nu}(n,s)$ is a $2 k$-dimensional plaquette
and $U_{\mu D}(n,s)$ is a plaquette
containing two link variables in the extra direction.
$\beta$ is the inverse gauge coupling for the plaquette $U_{\mu \nu}$ 
and $\beta_{s}$ is the one for the plaquette $U_{\mu D}$ .
In general , $\beta \neq \beta_{s}$ .
The fermion action $S_{F}$ on the Euclidean lattice,
in terms of the $2 k$-dimensional notation, 
is given by
\widetext
\begin{eqnarray}
 S_{F}& = & \frac{1}{2} \sum_{n \mu} \sum_s 
        \bar{\psi}_s(n) \gamma_\mu \left[ U_{s,\mu}(n) \psi_s(n + \mu)
        - U_{s,\mu}^{\dag}(n - \mu) \psi_s(n - \mu) \right]  \nonumber \\
& & \mbox{}+ \sum_n \sum_{s,t} \bar{\psi}_s(n) \left[ M_0 P_R 
        + M_0^{\dag} P_L \right] \psi_t(n)  \nonumber \\
& & \mbox{}+ \frac{1}{2} \sum_{n \mu} \sum_s
        \bar{\psi}_s(n) \left[ U_{s,\mu}(n) \psi_s(n + \mu)
        + U_{s,\mu}^{\dag}(n - \mu) \psi_s(n - \mu) -2 \psi_s(n) \right],
\label{eqn:fermion} 
\end{eqnarray}
\narrowtext
where $s , t$ are an extra coordinates , 
$P_{R/L} = \frac{1}{2} (1 \pm \gamma_{2k+1})$ , 
\begin{equation}
\left\{ 
\begin{array}{l}
 ( M_0 )_{s,t} = U_{s,D}(n) \delta_{s + 1 , t} - a(s) \delta_{s,t} \\
 ( M_0^{\dag} )_{s,t} = U_{s - 1 , D}^{\dag}(n) \delta_{s - 1 , t} - a(s) 
\delta_{s,t} . 
\end{array}
\right.
\end{equation}
Here $U_{s,\mu}(n) , U_{s,D}(n)$ ($D=2k+1$) are link variables 
connecting a site $(n,s)$ to $(n+\mu,s)$ or $(n,s+1)$, respectively.
Because of a periodic boundary condition in the extra dimension , 
$s , t$ run from $-L_{s}$ to $L_{s} - 1$ , 
and $a(s)$ is given by 
\begin{eqnarray}
  a(s) & = & 1 - m_0 \, {\rm{sign}}
\left[( s + \frac{1}{2} ) \, {\rm{sign}}( L_{s} - s - \frac{1}{2} ) \right] \nonumber \\
 & = & \left\{
\begin{array}{ll}
1 - m_0  & \mbox{}  ( - \frac{1}{2} < s < L_{s} - \frac{1}{2} ) \\
1 + m_0  & \mbox{}  ( - L_{s} - \frac{1}{2} < s < - \frac{1}{2} ) \, ,
\end{array}
\right.
\end{eqnarray}
where $m_{0}$ is the height of the domain-wall mass.
It is easy to check 
that the above fermionic action is identical to the one
in $2k+1$ dimensions, proposed by Kaplan\cite{kaplan,naraneu}.

In weak coupling limit of both $\beta$ and $\beta_{s}$ , 
it has been shown that at $0 < m_{0} < 1$
a desired chiral zero mode
appears on a domain wall ($s=0$ plane) without unwanted doublers.
Because of the periodic boundary condition in the extra dimension, 
however, a zero mode of the opposite chirality to the one on the domain wall 
appears on the anti-domain-wall $(s=L_{s} - 1)$. 
Overlap between two zero modes decreases exponentially at large $L_s$.
A free fermion propagator is easily calculated and
an effective action of a (2+1)- and (4+1)-dimensional model
including the gauge anomaly and the Chern-Simons term can be obtained for
smooth background gauge fields\cite{aokihirose,aokihirose1}.

The original Kaplan's domain-wall models in the 4+1 dimension, however, 
have not been investigated yet 
{\it non-perturbatively}, except $\beta_s =0$\cite{waveg,aoitnioshi,kornip}
and (2+1)-dimensional case\cite{aokinagai}.
Main question is whether the chiral zero mode survives 
in the presence of rough gauge fields mentioned in the introduction.
To answer this question we will analyze 
the fate of the chiral zero mode
in the weak coupling limit for $\beta$.
In this limit,
the gauge field action $S_{G}$ is reduced to
\begin{equation}
S_{G} = \beta_{s} \sum_{s} \sum_{n,\mu} 
\left\{ 1 - {\rm Re Tr} \left[ V(n,s) V^{\dag}(n+\mu,s) \right] \right\}, 
\label{eqn:gauge}
\end{equation}
where the link variable $U_{s,D}(n)$ in the extra direction
is regarded as a site variable $V(n,s)( = U_{s,D}(n))$.
This action is identical to the one of a $2 k$-dimensional spin model
and $s$ is regarded as an independent flavor.
The action eq.(\ref{eqn:gauge}) is invariant under
\begin{equation}
V(n,s) \longrightarrow g(s) V(n,s) g^{\dag}(s+1)\quad, \qquad
(g(s) \in G) , 
\label{eqn:symmetry}
\end{equation}
where $G$ is the gauge group of the original model.
Therefore the total symmetry of the model
is $G^{2 L_{s}}$, where $2L_s$, the size of the extra 
dimension, is regarded as the number of independent flavors.
We use this (reduced) model for our numerical investigation.
%
%
%
\subsection{Mean field approximation for fermion propagators}
When the dynamical gauge fields are added even on the extra dimension only, 
it is difficult to calculate the fermion propagator analytically.
Instead of calculating the fermion propagator {\it exactly}, 
we use a mean-field approximation 
to see an effect of the dynamical gauge field qualitatively.
The mean-field approximation we adopt is 
that the link variables are replaced as
\begin{equation}
V(n,s)\,\,\, \left(=U_{s,D}(n) \right) \,\,\, \longrightarrow z , 
\end{equation}
where $z$ is a $(n,s)$-independent constant.
From eq.(\ref{eqn:fermion}) 
the fermion action in a $2 k$-dimensional momentum space becomes
\widetext
\begin{equation}
S_{F} \rightarrow
\sum_{s,t,p} \bar{\psi}_{s}(-p) 
\left(\sum_{\mu} i \gamma_{\mu} \sin (p_{\mu}) \delta_{s,t}
+ \left[ M(z) P_{R} + M^{\dag}(z) P_{L} \right]_{s,t} \right) 
\psi_{t}(p), 
\end{equation}
\begin{equation}
(M(z))_{s,t} = (M_{0}(z))_{s,t} + \frac{\nabla(p)}{2} \delta_{s,t} \, ,
\,
(M^{\dag}(z))_{s,t} = (M^{\dag}_{0}(z))_{s,t} + \frac{\nabla(p)}{2} 
\delta_{s,t} \, , 
\,
\nabla(p) \equiv \sum_{\mu=1}^{D-1} 2 ( \cos p_{\mu} - 1 ) , \nonumber
\end{equation}
\begin{equation}
(M_{0}(z))_{s,t} = z\delta_{s+1,t}-a(s)\delta_{s,t} \quad , \qquad 
(M^{\dag}_{0}(z))_{s,t} = z\delta_{s-1,t}-a(s)\delta_{s,t} \quad .
\end{equation}
Following Refs.\cite{aokihirose,aokihirose1,naraneu}
,especially Ref.\cite{aokihirose1},
 it is easy to obtain
a mean field fermion propagator 
on a finite lattice with the periodic boundary condition:
\begin{eqnarray}
&& G(p)_{s,t} = \nonumber \\
&&\left[ \left\{ \left( 
        - i \sum_\mu \gamma_\mu \bar{p}_\mu + M(z) \right) 
G_L(p) \right\}_{s,t} P_L 
        \right. 
+ \left.  \left\{ \left(
         - i \sum_\mu \gamma_\mu \bar{p}_\mu + M^{\dag}(z) \right) 
G_R(p) \right\}_{s,t} P_R \right] , 
\label{eqn:propagator}
\end{eqnarray}
\narrowtext
\begin{equation}
 G_L(p) = \frac{\displaystyle 1}{\displaystyle {\bar{p}^2 + M^{\dag}(z) M(z)}} 
\quad , \quad 
 G_R(p) = \frac{\displaystyle 1}{\displaystyle {\bar{p}^2 + M(z) M^{\dag}(z)}} \quad , 
\label{eqn:formal}
\end{equation}
with $\bar{p}_{\mu} \equiv \sin (p_{\mu})$.
For large $L_{s}$ where we neglect terms of $O(e^{-c L_{s}})$ with $c > 0$, 
$G_{L}$ and $G_{R}$ are given by
\widetext
\begin{equation}
 \left[ G_L(p) \right]_{s,t} = \left\{ 
\begin{array}{ll}B e^{-\alpha_{+} |s - t|} 
        + \left( A_L - B \right) e^{- \alpha_{+}(s + t)} 
        + \left( A_R - B \right) e^{- \alpha_{+}(2L_{s} - s - t)} ,
 & (s , t \geq 0)  \\
A_L e^{- \alpha_{+} s + \alpha_{-} t} 
        + A_R e^{- \alpha_{+}(L_{s} - s) - \alpha_{-}(L_{s} + t)} ,
 & (s \geq 0 , t \leq 0) \\
A_L e^{ \alpha_{-} s - \alpha_{+} t}
        + A_R e^{- \alpha_{-}(L_{s} + s ) - \alpha_{+}(L_{s} - t)} ,
 & (s \leq 0 , t \geq 0) \\
C e^{-\alpha_{-} |s - t|}
        + \left( A_L - C \right) e^{ \alpha_{-}(s + t)}
        + \left( A_R - C \right) e^{- \alpha_{-}(2L_{s} + s + t)},
 & (s , t \leq 0)
\end{array}
\right.
\label{eqn:GL}
\end{equation}
\begin{equation}
 \left[ G_R(p) \right]_{s,t} = \left\{
\begin{array}{ll}
B e^{-\alpha_{+} |s - t|}
        + \left( A_R - B \right) e^{- \alpha_{+}(s + t + 2)}
        + \left( A_L - B \right) e^{- \alpha_{+}(2L_{s} - s - t -2)} ,
 & (s , t \geq -1)  \\
A_R e^{- \alpha_{+}(s + 1) + \alpha_{-} (t + 1)}
        + A_L e^{- \alpha_{+}(L_{s} - s -1) - \alpha_{-}(L_{s} + t +1)} ,
 & (s \geq -1 , t \leq -1) \\
A_R e^{ \alpha_{-} (s + 1) - \alpha_{+} (t + 1)}
        + A_L e^{- \alpha_{-}(L_{s} + s + 1) - \alpha_{+}(L_{s} - t -1)} , 
 & (s \leq -1 , t \geq -1) \\
C e^{-\alpha_{-} |s - t|}
        + \left( A_R - C \right) e^{ \alpha_{-}(s + t +2)}
        + \left( A_L - C \right) e^{- \alpha_{-}(2L_{s} + s + t +2)} , 
 & (s , t \leq -1)
\end{array}
\right.
\label{eqn:GR}
\end{equation}
\narrowtext
where 
\begin{eqnarray}
&& a_{\pm} = z ( 1 - \frac{\nabla(p)}{2} \mp m_{0} ) = z b_{\pm} , \\
&& \alpha_{\pm} = {\rm{arccosh}} 
 \left[ \frac{\bar{p}^{2} + z^{2} + b_{\pm}^2}{2 z b_{\pm}} \right] , \\
&& A_L = \frac{1}{a_{+} e^{\alpha_{+}} - a_{-} e^{- \alpha_{-}}}, \,
   A_R = \frac{1}{a_{-} e^{\alpha_{-}} - a_{+} e^{- \alpha_{+}}}, \\
&& B = \frac{1}{2 a_{+} \sinh \alpha_{+}} \quad , \quad 
   C = \frac{1}{2 a_{-} \sinh \alpha_{-}} . 
\end{eqnarray}
Like a free fermion theory,
the terms of $A_{R} , B$ and $C$ 
have no singularity for all $z$ as $p\rightarrow 0$.
A behavior of $A_{L}$ is, however, different.
As $p \rightarrow 0$ $A_{L}$ behaves as
\begin{equation}
A_{L}  \rightarrow \left\{ 
\begin{array}{lcl}
\frac{\displaystyle 1}{\displaystyle [(1 - m_{0})^{2}-z^2] + O(p^{2})}
& , & (0 < m_{0} < 1 - z) \\
\\
\frac{\displaystyle 4m_{0}^{2}-[(z^{2}-1)-m_{0}^{2}]^{2}}
{\displaystyle 4m_{0}z^{2}p^{2}}& , & (1 - z < m_{0} < 1) .
\end{array}
\right.
\end{equation}
A critical value of the domain-wall mass 
that separates a region with a zero mode 
and a region without zero modes 
is $m_{0}^c = 1 - z$.
Since $A_{L}$ term dominates for $1 - z  < m_{0} < 1$
in the $G_{L}$ (eq.(\ref{eqn:GL})) and $G_{R}$ (eq.(\ref{eqn:GR})), 
a right-handed zero mode appears in the $s=0$ plane , 
and a left-handed zero mode in the $s=L_{s}-1$ plane.
For $0 < m_{0} < 1 - z$ 
the right- and left-handed fermions are massive in all $s$-planes. 
Since the terms of $A_{L}  (A_{R}) $ and $B (C)$ are almost same value
in this region of $m_{0}$, 
a translational invariant term dominates
in $G_{L}$ and $G_{R}$ in the positive (negative) $s$-layer
, so that the spectrum becomes vector-like.

If $z \rightarrow 1$ , the model becomes a free theory.
The propagator obtained in this section agrees with the one 
obtained in Ref.\cite{aokihirose}.
In the opposite limit that $z \rightarrow 0$ , 
since there is no hopping term to the neighboring layers, 
this model becomes the one analyzed in Ref.\cite{aoitnioshi} 
in the case of the strong coupling limit $\beta_{s} = 0$ , 
and in Ref.\cite{kornip} , 
in the case that $z$ is identified to the vacuum expectation value of the link 
variables.
This consideration suggests that 
the region where the zero modes exist 
become smaller and smaller as $z$ $(1-z < m_{0} < 1)$ approaches zero.

\begin{figure}[t]
\centerline{\epsfxsize=12cm \epsfbox{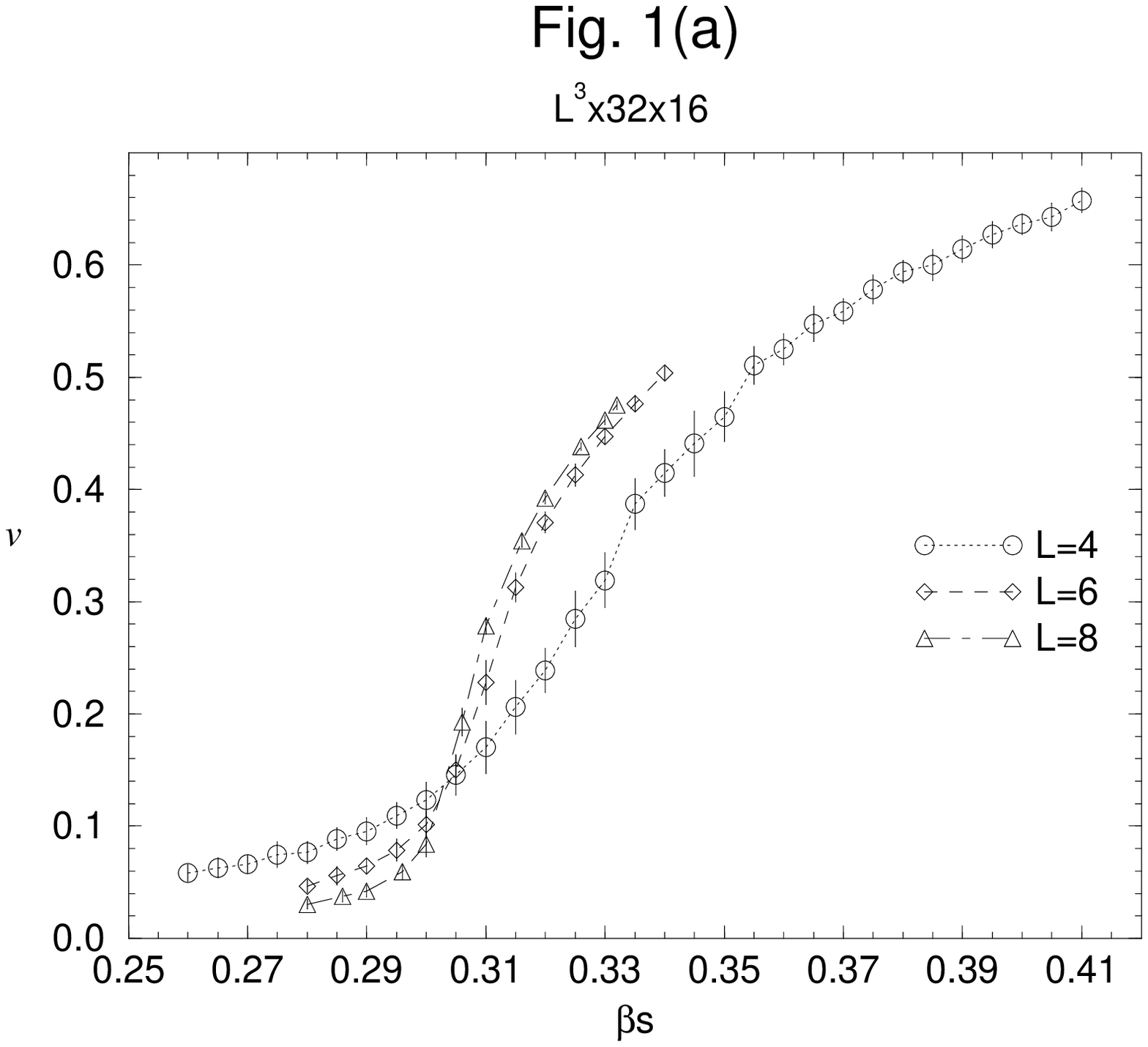}}
\centerline{\epsfxsize=12cm \epsfbox{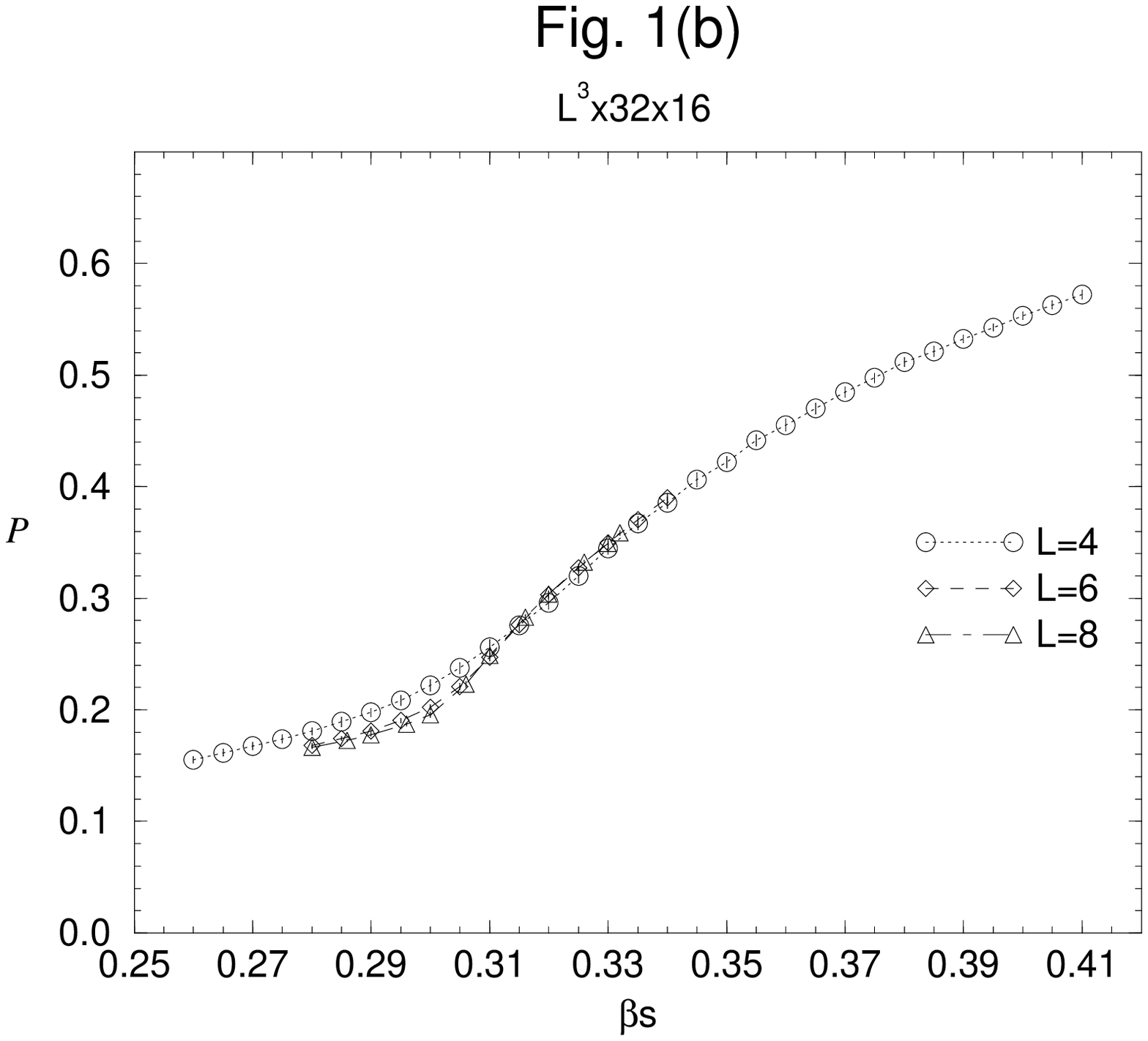}}
\caption{
(a) Vacuum expectation value of link variables $v$
and 
(b) Vacuum expectation value of plaquette $w$ 
on $L^3 \times 32 \times 16$ lattices 
with $L=$4(circles) , 6(diamonds) , 8(triangles)
as a function of $\beta_s$.}
\label{vev}
\end{figure}

What corresponds to $z$ ?
Boundary conditions which $z$ satisfies are
$z=1$ at $\beta_s=\infty$ and $z=0$ at $\beta_s=0$.
As explained before 
the gauge field action of our model 
is identical to that of the U(1) spin system
in 4-dimensions. 
So the most naive candidate\cite{kornip} is 
\begin{equation}
z= \langle V(n,s) \rangle ,
\label{eqn:orderp}
\end{equation}
which is not invariant under the symmetry (\ref{eqn:symmetry}).
In our simulations on $L^3 \times L_{4} \times 2 L_{s}$ lattices,
as an order parameter,
we take a vacuum expectation value of link variable
calculated with rotational technique:
\begin{equation}
v = \langle \,\, \frac{1}{2 L_s} \sum_{s}
 \left| \frac{1}{L_4 L^{3} } \sum_{n} V(n,s) \right| 
\,\,  \rangle
\quad . 
\label{eqn:order}
\end{equation}
Although $v$ defined in eq.(\ref{eqn:order}) is always non-zero on finite lattices,
it becomes zero in the symmetric phase but stays non-zero
in the broken phase in the infinite volume limit.
Fig. \ref{vev}(a) shows that,
even on finite lattices,
$v$ behaves as if it was an order parameter:
it is very small and decreases as the volume increases
in the would-be symmetric phase.
If the identification that $z=v$ is true, 
zero modes disappear in the symmetric phase, where $v = 0$.
The other choice, which is invariant under (\ref{eqn:symmetry}), is
the vacuum expectation value of plaquette:
\begin{equation}
z^2 = \langle {\rm Tr Re} \{V(n,s) V^\dagger(n+\mu,s)\} \rangle .
\label{energy}
\end{equation}
In this identification
the zero modes always exist in both phases, since
$\langle {\rm Tr Re}\{ V(n,s) V^\dagger(n+\mu,s)\} \rangle$ 
is non-zero for all $\beta_s$ except $\beta_s=0$ 
and is insensitive to which phase we are in,
as shown in Fig. \ref{vev}(b).
%
%
%
%
\section{numerical study of (4+1)-dimensional U(1) model}
\subsection{Method of numerical calculations}
In this section
we numerically study the domain-wall model in 4+1 dimensions
with a U(1) dynamical gauge field in the extra dimension.
As seen from eq.(\ref{eqn:gauge}) , 
the gauge field action can be identified with a 4-dimensional U(1) spin 
model (with $2 L_{s}$ copies).

Our numerical simulation has been carried out by the quenched approximation.
Configurations of U(1) dynamical gauge field are generated
and fermion propagators are calculated on these configurations.
The obtained fermion propagators are gauge non-invariant in general 
under the symmetry (\ref{eqn:symmetry}).
The fermion propagator $G(p)_{s,t}$ becomes ``invariant'' 
if and only if $s=t$.
Thus, we take the $s-s$ layer as propagating plane 
($\approx$ ``physical space''), 
and investigate the behavior of the fermion propagator in this layer. 

To study the fermion spectrum, 
we first extract $G_{L}$ and $G_{R}$, defined in eq.(\ref{eqn:propagator}),
from the fermion propagator.
Assuming eq.(\ref{eqn:formal}),
we then obtain corresponding ``fermion masses''
from $G_{L}^{-1}(p)$ and $G_{R}^{-1}(p)$ by fitting them linearly 
in $\bar{p}^{2}$ as follows.
\begin{eqnarray}
G_{L}^{-1} &=& \bar{p}^{2} + M^{\dag} M \rightarrow m_{f}^{2} ({\rm{Right}})
\quad , \quad (p \rightarrow 0) , 
\label{eqn:Lanalysis} \\
G_{R}^{-1} &=& \bar{p}^{2} + M M^{\dag} \rightarrow m_{f}^{2} ({\rm{Left}})
\quad , \quad (p \rightarrow 0) .
\label{eqn:Ranalysis} 
\end{eqnarray}
We take the following setup for 4-dimensional momenta.
A periodic boundary condition is taken for the 1st- 2nd- 3rd-direction
and the momenta in these directions are fixed on $p_{1}, p_2, p_3 = 0$.
An anti-periodic boundary condition is taken for the 4th-direction
and the momentum in this direction is variable,
$p_{4} = (2n+1) \pi / L_4 \, , \, n = - L_4 / 2 ,..., L_4 / 2 - 1$,
where $L_4$ is the number of site of 4th direction.
(Our numerical simulations have been performed always for $L_4 = 32$.)
%
%
\subsection{Simulation parameters}
From Fig. \ref{vev}(a) 
it is inferred that 
the system is in the symmetric phase at $\beta_s = 0.29$
and in the broken phase at $\beta_s=0.5$.
Our simulation is performed in the quenched approximation on
$L^{3}\times 32 \times 2 L_{s}$ lattices with $L = $4, 6, 8 and $L_{s}= 8$ 
at $\beta_s = $ 0.29 (symmetric phase) ,
where $L$ is a lattice size of 1st- 2nd- 3rd-direction, 
and on $L^{3}\times 32 \times 2 L_{s}$ lattices with $L = $4, 6, 8 and $L_s= 8$
at $\beta_s = $ 0.5 (broken phase).
The coordinate $s$ in the extra dimension runs $-8 \leq s \leq 7$. 
Gauge configurations are generated by the 5-hit Metropolis algorithm.
For the thermalization first 5000 sweeps are discarded.

The fermion propagators are calculated by the conjugate gradient method
on 50 configurations separated by 100 sweeps.
We take the domain-wall mass 
$m_{0} =$ 0.7 , 0.8 , 0.9 , 0.95 , 0.99 at $\beta_{s} =$ 0.29 and 
$m_{0} =$ 0.1 , 0.2 , 0.3 , 0.4 , 0.5 , 0.6 , 0.9  at $\beta_{s} =$ 0.5.
As mentioned before,
the boundary conditions in 1st- 2nd- 3rd- and 5th-directions
are periodic and the one in 4th-direction is anti-periodic.
Wilson parameter $r$ has been set to $r = 1$.
The fermion propagators have been investigated mainly at $s =$ 0 , $-1$.
These $s$ are the layers where we put sources.
The layer at $s=0$ is the domain wall.
Errors are all estimated by the jack-knife method with unit bin size.
%
%
\subsection{Fermion spectrum in the broken phase}
\begin{figure}
\centerline{\epsfxsize=12cm \epsfbox{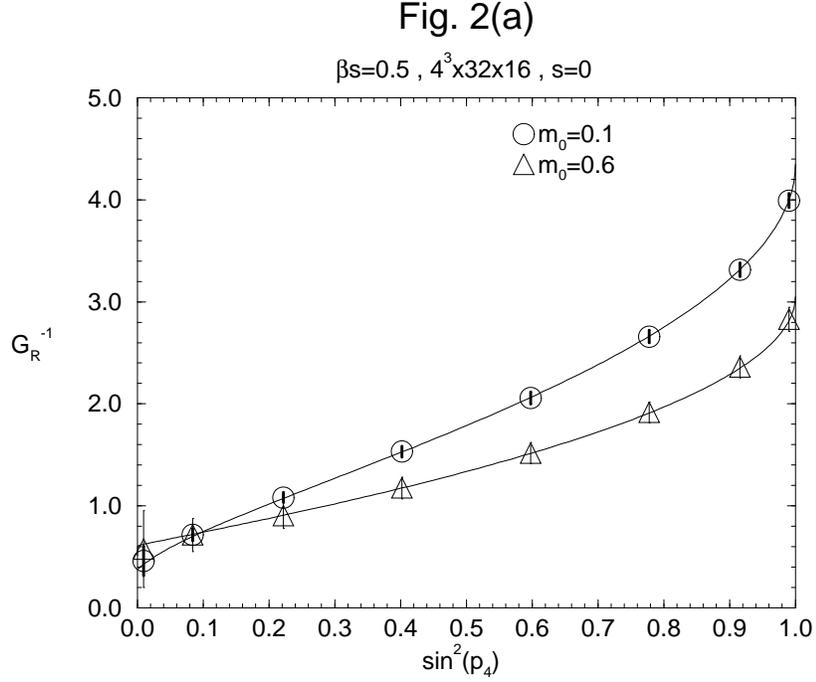}}
\centerline{\epsfxsize=12cm \epsfbox{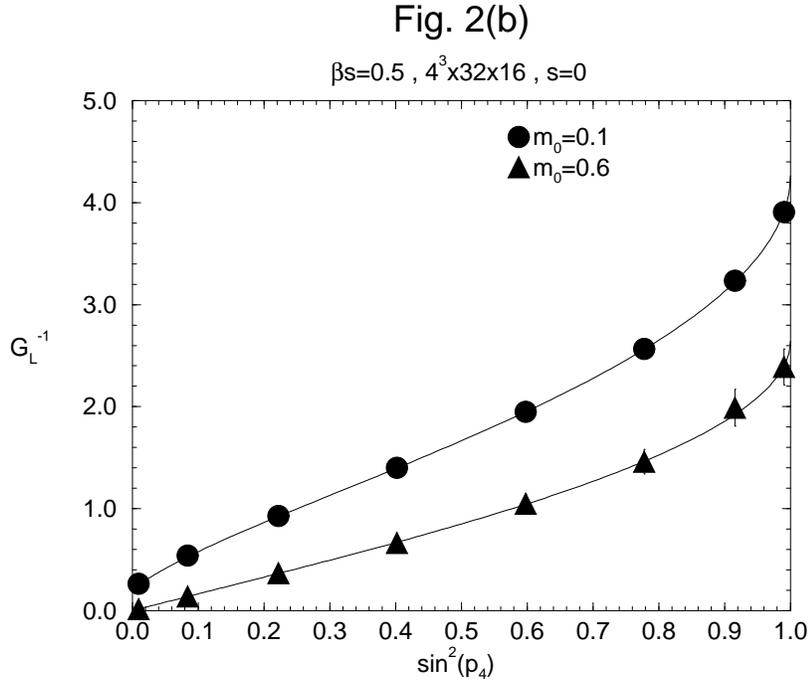}}
\caption{
(a) $[G_{R}]_{0,0}^{-1}$ as a function of $\sin^{2}(p_{4})$
and 
(b) $[G_{L}]_{0,0}^{-1}$ as a function of $\sin^{2}(p_{4})$
with $p_1 , p_2 , p_3 =0$
at $\beta_{s}=0.5$ on a $4^3 \times 32 \times 16$ lattice,
at $m_0$=0.1 (solid circles) and 0.6 (solid triangles).
Solid  lines of both figures stand for the ones 
obtained from the mean-field propagator with
the fitted parameter $z$. }
\label{invprop}
\end{figure}

\begin{figure}
\centerline{\epsfxsize=12cm \epsfbox{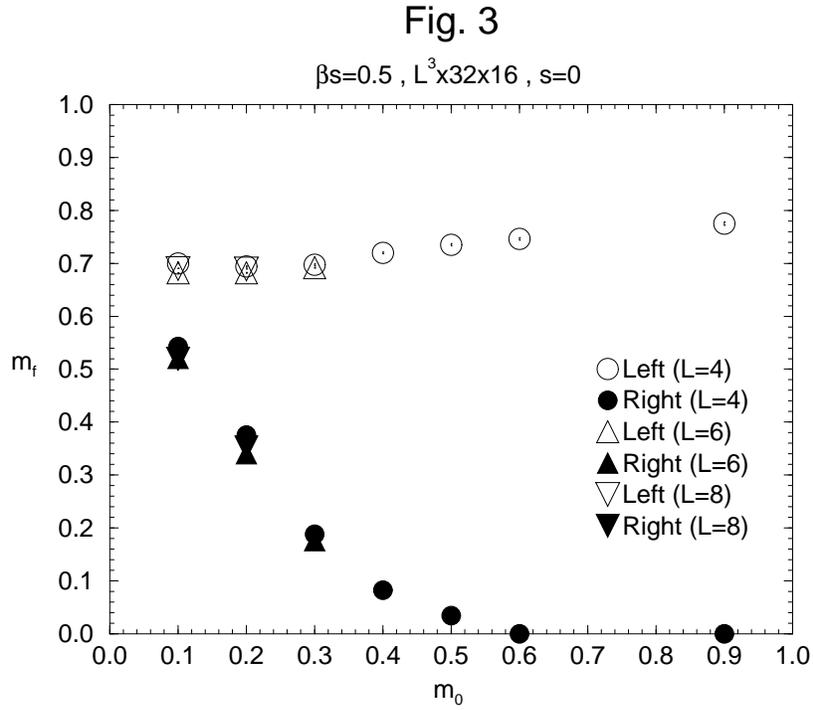}}
\caption{
$m_{f}$ vs. $m_{0}$ 
at $\beta_{s} = 0.5$ on a $L^3 \times 32 \times 16$ lattices 
with $L=$4(circles) , 6 (up triangles) and 8 (down triangles)
in the case of putting a source on the domain wall at $s=0$,
for the right-handed fermion (solid symbols) and the left-handed
fermion (open symbols).}
\label{mfs0}
\end{figure}

The system is in broken phase at $\beta_{s} =$ 0.5.
We first consider the fermion spectrum on the layer at $s=0$.
Let us show Fig. \ref{invprop}, which
is a plot of the $G_{L}^{-1}$ and $G_{R}^{-1}$ as a function of
$\bar{p}_4^{2} \equiv \sin^{2} (p_{4})$ at $m_0$ =0.1 and 0.6.
(Note we always set $p_1, p_2, p_3=0$.)
In the limit $p_{4} \rightarrow 0$, 
$G_{R}^{-1}$ remains non-zero at both $m_{0}$,
while $G_{L}^{-1}$ vanishes at $m_{0} = 0.6$.
We obtain the value of $m_f^2$, which can be regarded as the mass 
square in $4$-dimensional world,
by the linear fit in $\bar{p}_4^{2}$ near $\bar{p}_4^{2}=0$,
and plot $m_f$ as a function of $m_{0}$ in Fig. \ref{mfs0}.
The mass of right-handed fermion, obtained from $G_{L}^{-1}$,
becomes very small (less than 0.1) at $m_{0}$ larger than $0.35$, so
we conclude that the critical value is $m_{0}^{c} \sim 0.35$.
Whenever the domain-wall mass is larger than this value , 
this model produces the right-handed chiral zero mode on the domain wall
at $s=0$. 
From these results above
we conclude
that the domain-wall model with the dynamical gauge field
on the extra dimension ({\it {i.e.}}
the weak coupling limit of the original  model)
can survive the chiral zero mode on the domain-wall,
at least deep in the broken phase.

Also in Fig. \ref{invprop},
solid lines stand for the inverse propagators
obtained from the mean-field propagator 
with appropriately tuned parameter $z$,
and the lines show that
the behavior of the fermion propagator is
well described by the mean-field propagator.

An important question here is 
what a fermion spectrum is in the scaling limit.
If the fermion spectrum stays chiral in the limit,
it should stay chiral also in the symmetric phase,
since the phase transition is continuous.
This means that, in order to determine the fermion spectrum 
in the scaling limit even from the broken phase, 
we have to know the spectrum in the symmetric phase. 
Therefore, from the knowledge of the fermion spectrum 
obtained in the broken phase so far,
we cannot draw any conclusions on the fermion spectrum, 
chiral or vector-like, in the scaling limit.
%
%
\subsection{Fermion spectrum in symmetric phase}
The system is in the symmetric phase at $\beta_{s} = 0.29$.
The fermion propagator is analyzed in the same way as in the broken phase.

\begin{figure}
\centerline{\epsfxsize=12cm \epsfbox{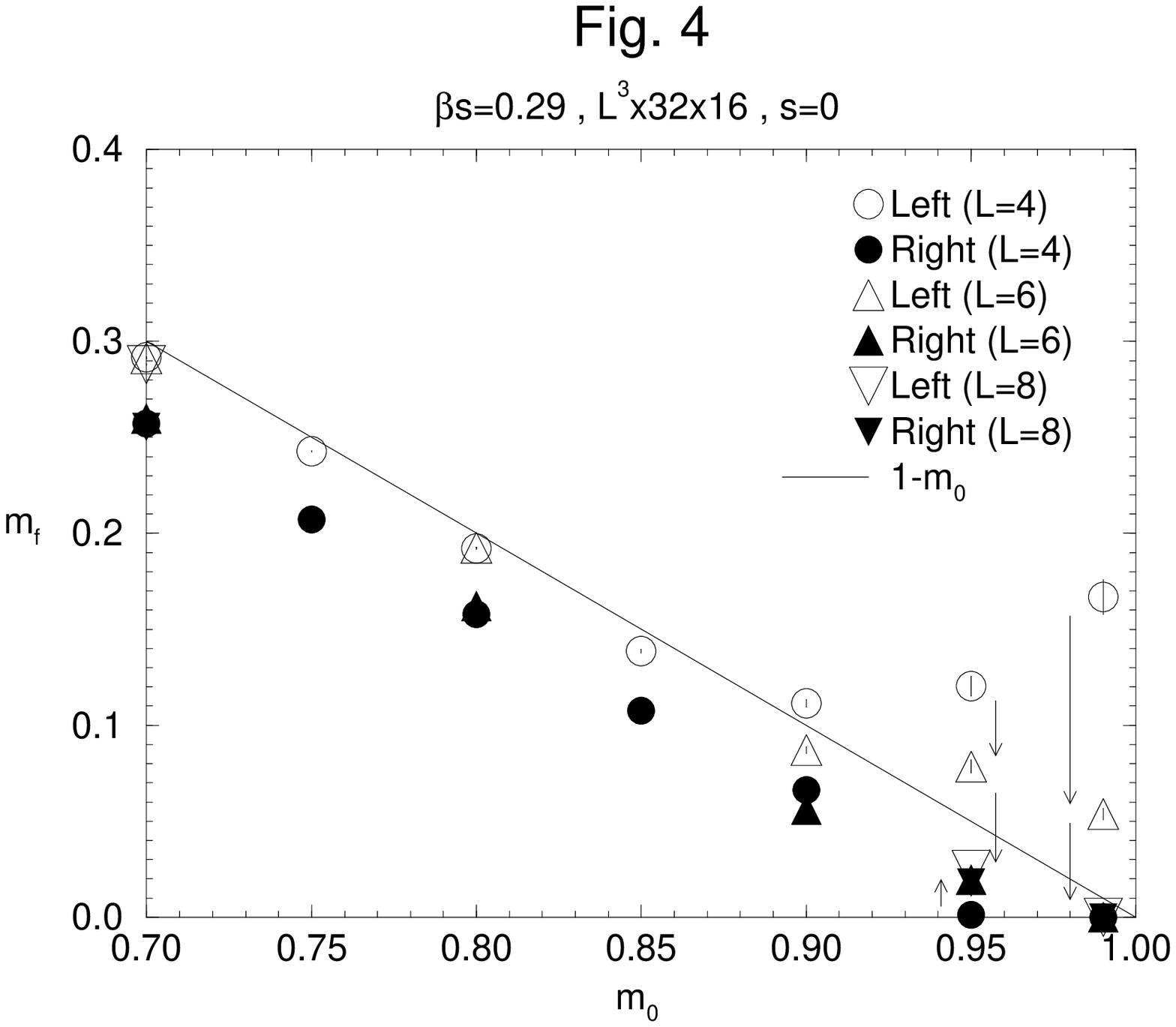}}
\caption{
$m_{f}$ vs. $m_{0}$
at $\beta_{s} = 0.29$ on $L^3 \times 32 \times 16$ lattices
with $L=$4 (circles) , 6 (up triangles) and 8 (down triangles) 
in the case of putting a source on the domain wall at $s=0$,
for the right-handed fermion (solid symbols) and the left-handed
fermion (open symbols).
Solid line corresponds to $1-m_0$.}
\label{symmfs0}
\end{figure}

\begin{figure}
\centerline{\epsfxsize=12cm \epsfbox{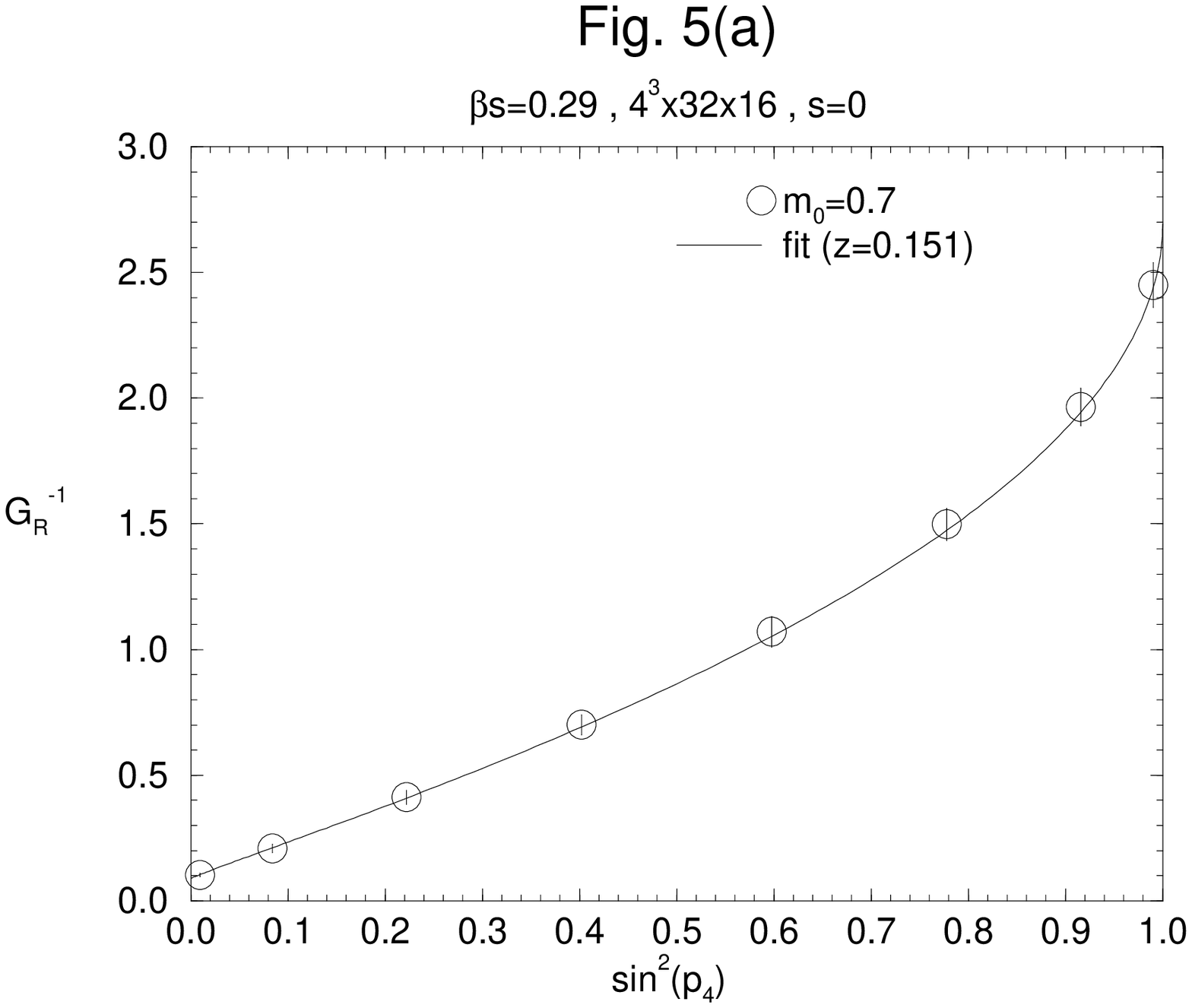}}
\centerline{\epsfxsize=12cm \epsfbox{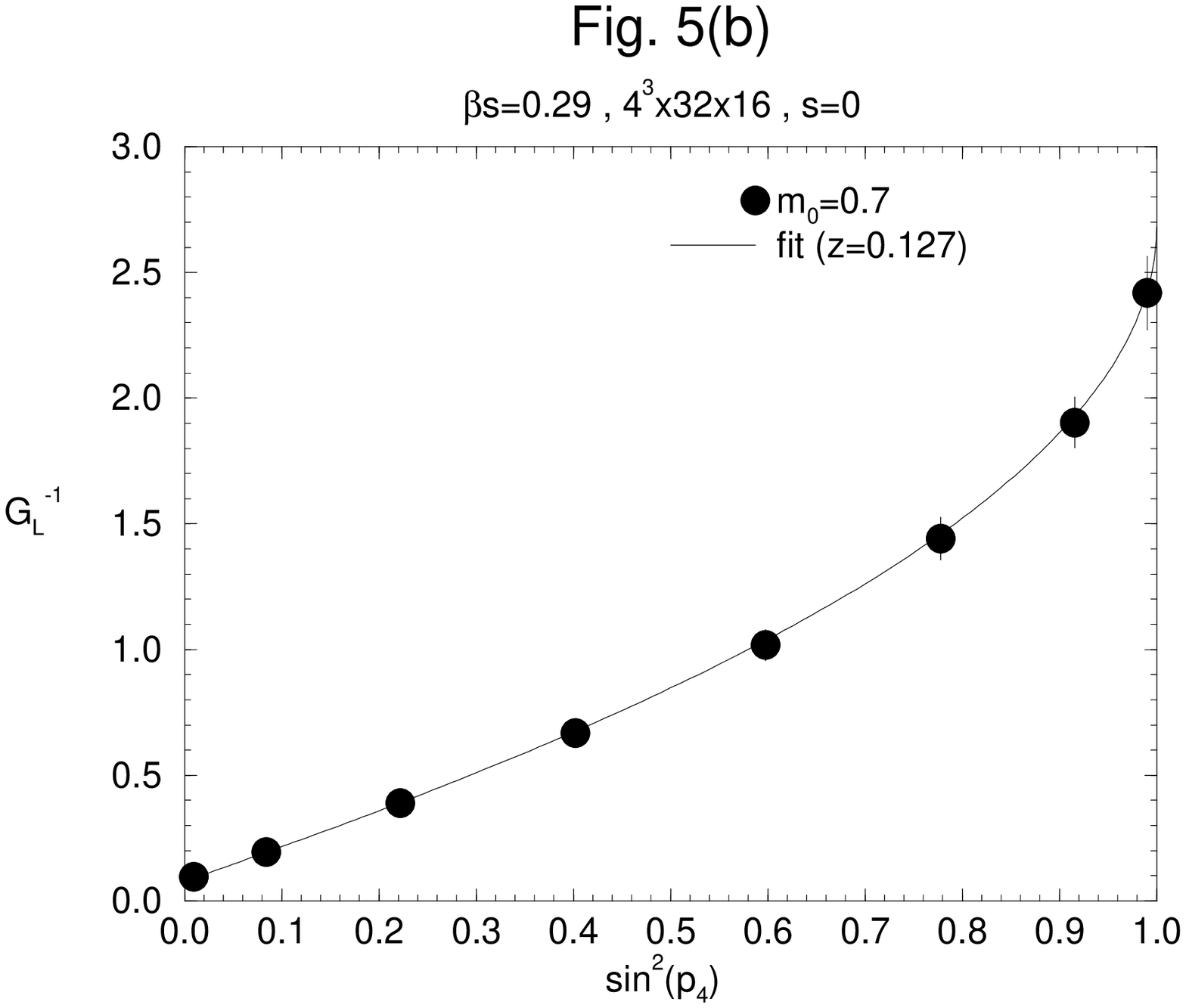}}
\caption{
(a) $[G_{R}]_{0,0}^{-1}$ as a function of $\sin^{2}(p_{4})$
and 
(b) $[G_{L}]_{0,0}^{-1}$ as a function of $\sin^{2}(p_{4})$
with $p_1 , p_2 , p_3 =0$
at $\beta_{s}=0.29$ on a $4^3 \times 32 \times 16$ lattice,
at $m_0$=0.7.
Solid lines of both figures stand for the ones 
obtained from the mean-field propagator with
the fitted parameter $z$. }
\label{invprops}
\end{figure}

\begin{figure}
\centerline{\epsfxsize=12cm \epsfbox{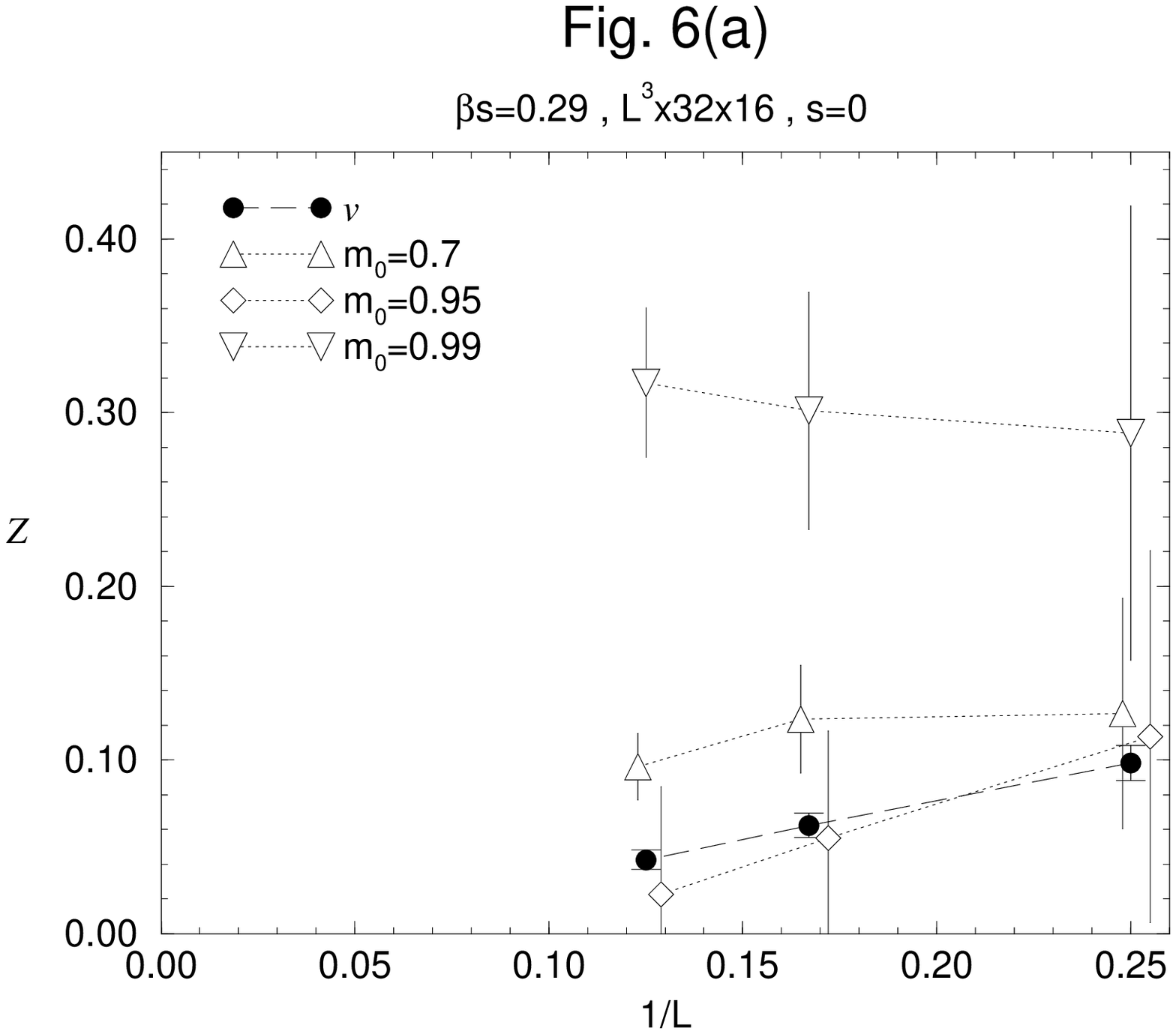}}
\centerline{\epsfxsize=12cm \epsfbox{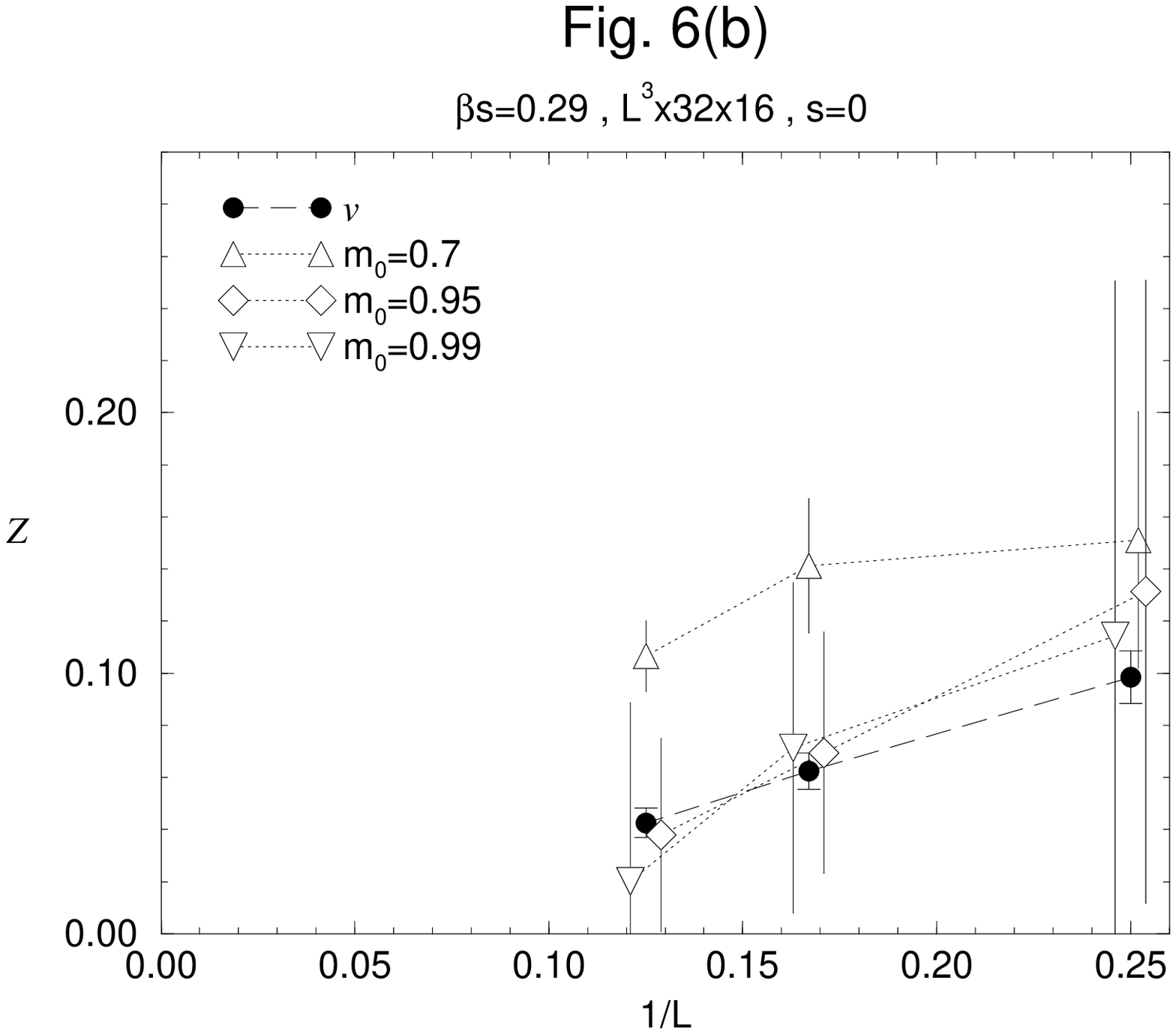}}
\caption{
(a) $z$(right-handed) and $v$ vs. $1/L$ at $\beta_s=0.29$
and 
(b) $z$(left-handed) and $v$ vs. $1/L$ at $\beta_s=0.29$. 
on $L^{3} \times 32 \times 16$ lattices
with $L=$4, 6 and 8 at
$m_0 = 0.7$ (up triangles), $0.95$ (diamonds) and $0.99$ (down triangles),
in the case of putting a source on the domain wall at $s=0$. 
Solid circles stand for the vacuum expectation value of link variable
(:the order parameter).}
\label{volumedep}
\end{figure}

In Fig. \ref{symmfs0}, 
we have plotted mass $m_f$ of the right- and left-handed modes
at $s=0$ as a function of $m_0$.
On a $4^3 \times 32 \times 16$ lattice,
it seems that the right-handed fermion becomes massless 
at $m_{0}$ larger than $0.95$, 
while the left-handed fermion stays massive at all $m_{0}$,
so that the fermion spectrum on the domain wall is chiral.
However as the lattice sizes become larger, 
for example $6^3 \times 32 \times 16 $ and $ 8^3 \times 32 \times 16$,
mass differences between the right- and the left-handed modes
become smaller.
This suggests that the fermion spectrum becomes
vector-like in the infinite volume limit.
From this data alone, however, we cannot exclude a possibility
that the critical mass $m_{0}^c$ is very close to 1.0,
since the fermion mass near $m_0=1.0$ is very small.
In order to make a definite conclusion
on the absence of chiral zero modes 
in the symmetric phase,
we try to fit $G_L^{-1}$ and $G_R^{-1}$ at given $m_0$
using the form of mean-field propagator, eqs.(\ref{eqn:GL})
and (\ref{eqn:GR}),
with the fitting parameter $z$.
We show the quality of this fit in Figs. \ref{invprops} (a) and (b).
These figures show that 
the fermion propagator is well described by the mean-field propagator
if the fitting parameter $z$ is chosen 
such that $\chi^2$ is minimized.
We then have plotted $z$ obtained by the fit
as a function of $1/L$ in Figs. \ref{volumedep} (a) and (b),
where $L$'s take 4, 6, 8 on the $L^3 \times 32 \times 16$ lattices. 
The values of $z$'s  are almost independent of $m_0$
at the each $1/L$
except the right-handed modes 
at $m_0=0.99$.
The solid circles represent the order parameter $v$
defined in eq.(\ref{eqn:order}).
The behaviors of $z$'s at different $m_0$
are almost identical each other
and are very similar to that of $v$
except the right-handed ones at $m_0=0.99$.
We think that the observed deviation of $z$(right-handed) at $m_0=0.99$ from those 
at other $m_0$'s is not a real effect but a statistical fluctuation,
since the $\chi^2$ for the fit at $m_0=0.99$
is almost flat in the region between $z=0$ and $z \sim 0.3$.
Ignoring this data at $m_0=0.99$
we see that $z$ can be identified with $v$,
and therefore $z$ becomes zero 
as the lattice size goes to infinity
since the order parameter $v$ defined in eq.(\ref{eqn:order}) becomes zero 
in the infinite volume limit.
This analysis leads us to a conclusion that
the fermion spectrum in the symmetric phase 
is {\it vector-like}.

\begin{figure}
\centerline{\epsfxsize=12cm \epsfbox{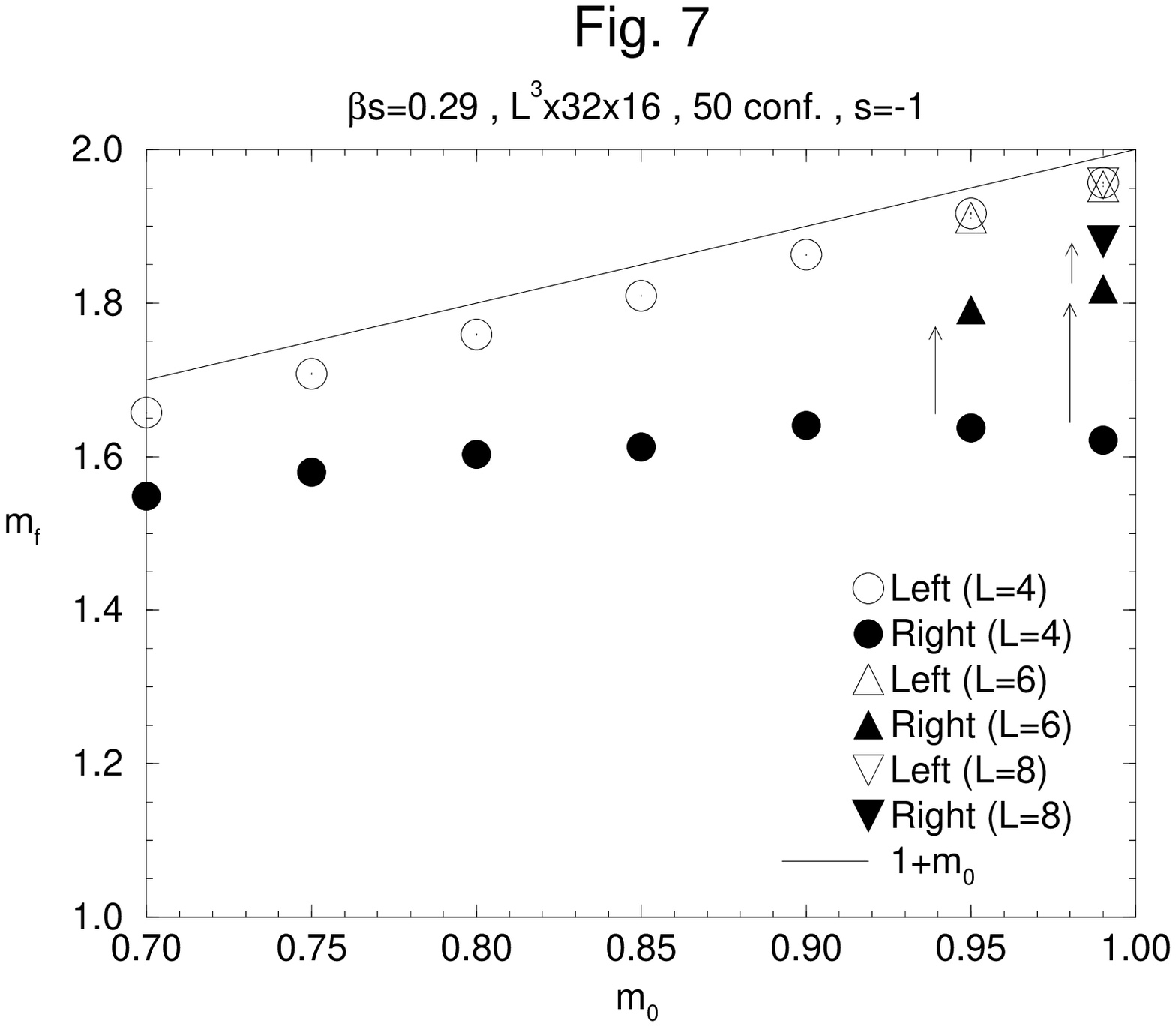}}
\caption{
$m_{f}$ vs. $m_{0}$
at $\beta_{s} = 0.29$ on $L^3 \times 32 \times 16$ lattices
with $L=$4 (circles) , 6 (up triangles) and 8 (down triangles)
in the case of putting a source at $s=-1$,
for the right-handed fermion (solid symbols) and the left-handed
fermion (open symbols).
Solid line corresponds to $1+m_0$.}
\label{symmfsm1}
\end{figure}

Fig. \ref{symmfs0} also shows that
the fermion masses tend to approach the line $m_f=1-m_0$,
which is the value of the mass for free Wilson fermion at a
positive $s$-layer, as the lattice volume increases.
Furthermore in Fig. \ref{symmfsm1} as a function of $m_0$,
we have also plotted $m_f$ at $s=-1$ (a negative $s$-layer).
As the volume increases
the mass difference between right-handed and left-handed fermions
becomes smaller and both masses seem to approach to the line $m_f=1+m_0$,
the value of the mass for free Wilson fermion at a negative $s$-layer.
This data also supports our conclusion on the absence of chiral zero modes
in the symmetric phase.

In summary
our results of $m_f$ at both positive $s$- and negative $s$-layer
suggest that chiral zero modes in the symmetric phase
disappear in the infinite volume limit.
This concludes that the original Kaplan's model fails to describe
lattice chiral gauge theories in the symmetric phase.
\section{Conclusions and discussions}
Using the quenched approximation, 
we have performed the numerical study of the domain-wall model 
in 4+1 dimensions
with the U(1) dynamical gauge field on the extra dimension.
From this study we obtain the following results.
In the broken phase of the gauge field,
there exists the critical value of the domain-wall mass 
separating the region with a chiral zero mode
and the region without it
in the same case of (2+1)-dimensional model.
At the domain-wall mass larger than its critical value,
a zero mode with one chirality 
exists on the domain wall.
In the symmetric phase, 
on the other hand, 
our data on $L^3 \times 32 \times 16$ with $L=$4 , 6 , 8 lattices 
suggest the absence of chiral zero modes
in the infinite volume limit, though
the chiral zero mode seems to exist on finite lattices.
We also found that fermion propagators obtained through
numerical simulations on finite lattices
are well described by the mean-field propagator
with $z \simeq v$.
Since the existence of chiral zero modes in the symmetric phase 
is essential for the success of the original domain-wall model, 
our results for the (4+1)-dimensional model indicate
that the original domain-wall model 
cannot work as lattice chiral gauge theories.

\begin{figure}
\centerline{\epsfxsize=12cm \epsfbox{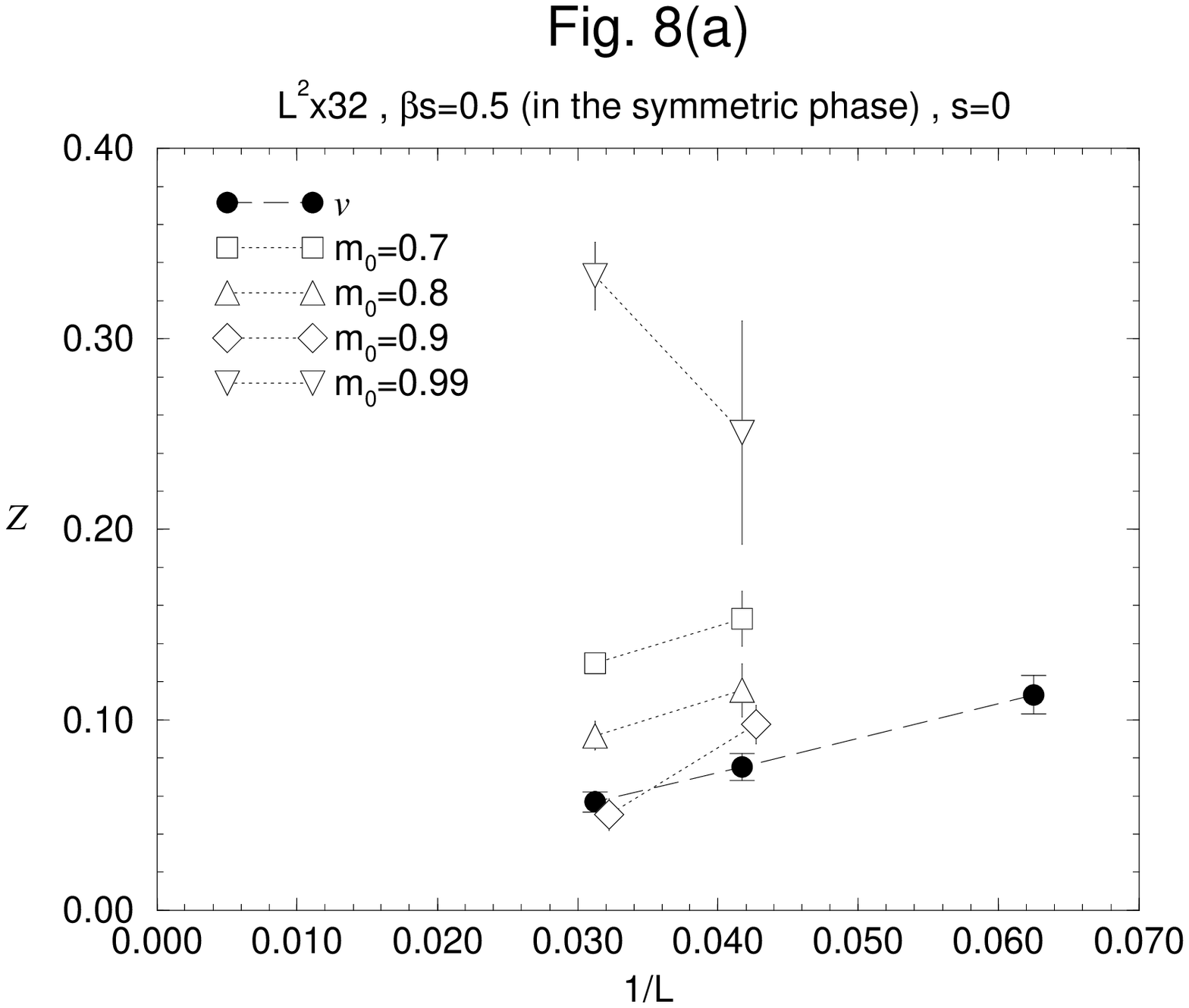}}
\centerline{\epsfxsize=12cm \epsfbox{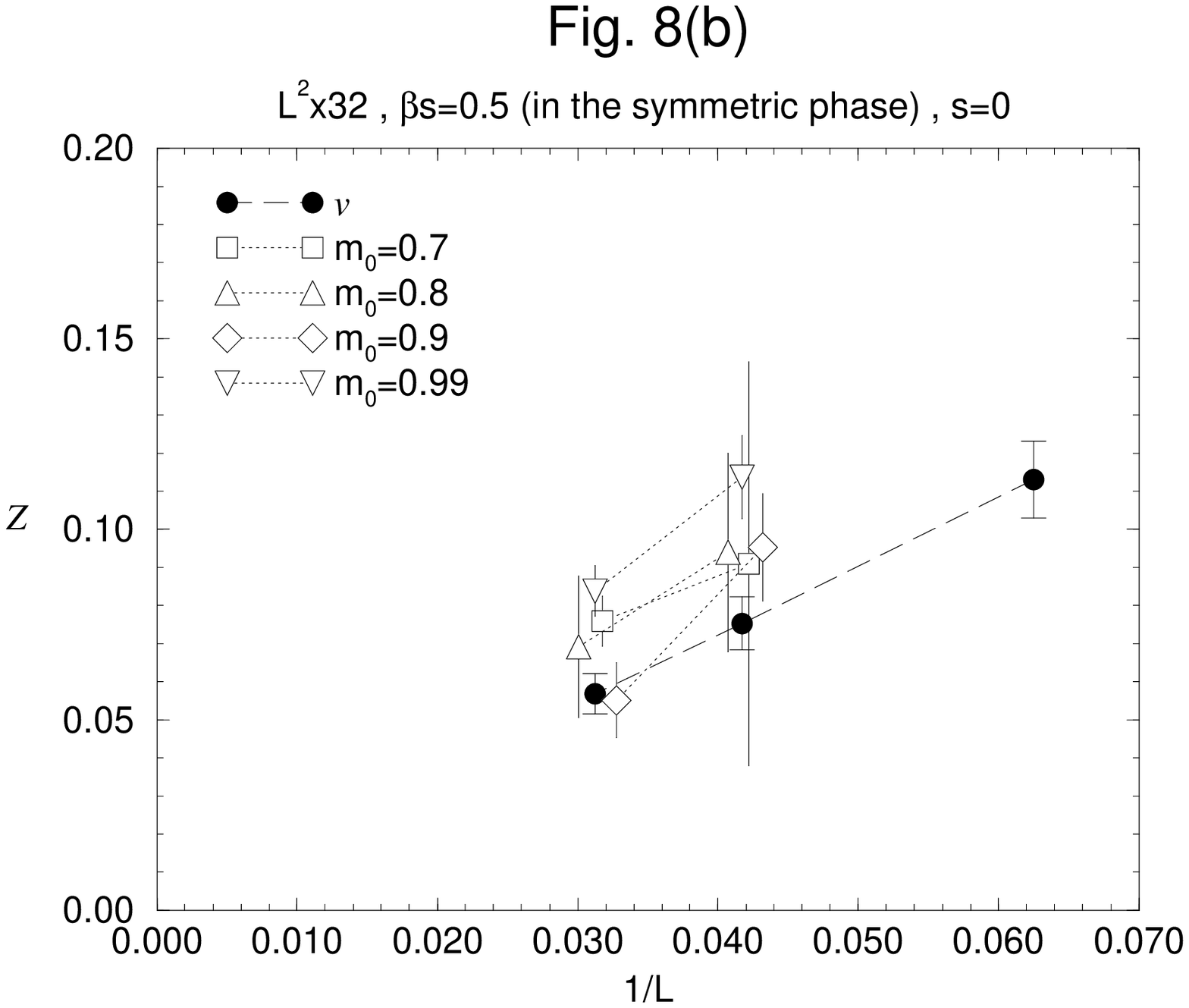}}
\caption{
(a) $z$(right-handed) and $v$ vs. $1/L$ at $\beta_s=0.5$
and 
(b) $z$(left-handed) and $v$ vs. $1/L$ at $\beta_s=0.5$
in the symmetric phase
on $L^{2}\times 32$ lattices
with $L=$24 and 32
at $m_0 = 0.7$ (squares), $0.8$ (up triangles),
$0.9$ (diamonds) and $0.99$ (down triangles),
in the case of putting a source on the domain wall at $s=0$.
Solid circles stand for the vacuum expectation value of link variable
(:the order parameter).}
\label{z2d}
\end{figure}

In 2+1 dimensions
we could not conclude 
whether the fermionic zero mode exists or not
because of the difficulty for the simulation near $m_0 = 1$
and the absence of the order parameter in the infinite volume limit 
for the 2-dimensional U(1) model.
We try here to extract the parameter $z$ of the (2+1)-dimensional model,
by the method used for the (4+1)-dimensional model.
In Figs. \ref{z2d} (a) and (b), the parameter $z$ is plotted 
as a function of $1/L$. The behavior of $z$ is almost identical to 
that of order parameter $v$ in eq.(\ref{eqn:order})
but not to the square root of plaquette in eq.(\ref{energy}).
This new analysis shows that
zero modes are absent in the symmetric phase
in the (2+1)-dimensional U(1) model as well as
the (4+1)-dimensional model.
Therefore we conclude that U(1) chiral gauge theories  can not be
constructed via an original domain-wall model,
regardless of its dimensionality.

One of the remaining question is whether the above negative conclusion
also holds for other gauge groups such as SU(2).
The answer is not so straightforward:
For example,  the 2-dimensional SU(2) spin model 
has a symmetric phase only.
We wonder whether chiral zero modes are absent
in the symmetric phase even if the gauge field
of the (2+1)-dimensional domain-wall model becomes ``smooth''
for large but finite $\beta_s$.
To answer this question
we are currently investigating
the (2+1)-dimensional SU(2) domain-wall model.
%
%
%
\section*{Acknowledgements}
Numerical calculations for the present work have been carried out
at Center for Computational Physics
and on VPP500/30 at Science Information Center,
both at University of Tsukuba. 
This work is supported 
in part by the Grants-in-Aid of the Ministry of Education
(Nos. 04NP0701, 08640349).
%
%
%
%

%
%
\newpage
%
%
\newpage

\newpage

\newpage

\newpage

\newpage

\newpage

\newpage

\end{document}